\documentclass[aps,%
prl,%
twocolumn,%
showpacs,%
preprintnumbers,%
superscriptaddress]{revtex4}
\usepackage{epsfig} 

\newcommand{\mysection}[1]{\paragraph*{#1.}}

\begin{document}

\author{Jan Nagler}
\affiliation{Institut f\"ur Theoretische Physik, Universit\"at Bremen, Otto-Hahn-Allee, 28334 Bremen, Germany}

\author{Jens Christian Claussen}
\email{claussen@theo-physik.uni-kiel.de}
\affiliation{Institut f\"ur Theoretische Physik und Astrophysik, Universit\"at Kiel, Leibnizstra{\ss}e 15, 24098 Kiel, Germany}

\title{$1/f^\alpha$ spectra in elementary cellular automata and fractal signals}
\date{October 19, 2004}
\newcommand{\mod}{{\rm{}mod}}
\newcommand{\oofs}{$1/f^{\alpha}$ spectra}

\begin{abstract}
We systematically compute the power spectra of the one-dimensional
elementary cellular automata introduced by Wolfram. On the one
hand our analysis reveals that one automaton displays $1/f$
spectra though considered as trivial, and on the other hand that
various automata classified as chaotic/complex display no $1/f$
spectra. We model the results generalizing the recently
investigated Sierpinski signal to a class of fractal signals 
that are tailored to produce $1/f^{\alpha}$ spectra. 
From the widespread occurrence of  (elementary) cellular automata 
patterns in chemistry, physics and computer sciences, there are 
various candidates to show spectra similar to our results.
\end{abstract}
\maketitle

In 1984 Wolfram introduced the so-called elementary cellular automata (ECA),
opening a field still being vividly active 20 years thereafter \cite{wolfram}.
Wolfram's more recent
popular book \cite{nks} has attracted
great attention, although the opinion of the work's merits are
divided among the scientific community \cite{nks_review}.
ECA are discussed extensively in the context of computationally irreducibility
of physical systems \cite{israeli2004},
e.g.\ it is proven that in the Turing sense \cite{herken1995}
rule 110 (being one of the possible 256 ECA) is an universal computer \cite{wolfram}.
Moreover, possible transformations
between difference equations and (E)CA have been investigated \cite{nobe2001}.
Among the numerous physical applications we mention here only
(kinetic phase transitions in) catalytic reaction-diffusion systems
\cite{ziff1986,dress84plath,%
hayase,claussen2004},
deterministic surface growth \cite{krug1988},
branching and annihilating random walks \cite{cardy1997} and
random boolean networks \cite{matache2004}.\\
\indent
It is important to
note that Wolfram's ECA are often studied for different boundary
conditions on a finite array. A particular boundary condition
(e.g. a periodic or an absorbing one) disturbs the {\em pure}
evolution of an ECA. As a result, some automata display complex
behavior, while other are simply periodic. Though there is no
algorithm for classifying a given elementary automaton, Wolfram
conjectured that ECA can be grouped
into four classes of complexity:\\
\indent
Class 1: Steady state,
class 2: Periodic or nested structures,
class 3: Random (chaotic) behavior,
class 4: Mixture of random and periodic behavior.

The first class represents automata that are (for almost all initial conditions) trivial
in the sense being static or finally evolve to the some steady state.
Those rules that belong to the second class produce simple periodic or
self-similar, i.e. fractal, structures.
The third class includes rules exhibiting random patterns, e.g. a particular rule (number 30)
is used to generate random numbers in {\em Mathematica}.
The fourth class is somehow a mixture of classes 2 and 3
generating the most complex structures.
For more rigorous classifications
we refer the reader to the literature \cite{barbosa2004,israeli2004}.

\indent
Since the coining paper of Bak, Tang, and Wiesenfeld \cite{btw},
there has been considerable interest in the long-time
behavior of cellular automata, especially for occurrence
of long range correlations, and correspondingly for
power spectra exhibiting a power law decay $S(f)\sim f^{-\alpha}$
with $\alpha \approx 1.0$.
Despite the abundance in nature, systems exhibiting spectra with
exponents near to 1 are poorly understood.
While the mechanisms generating
$1/f^\alpha$ spectra may be
substantially different from each other,
some models 
and the observed $1/f^\alpha$ power laws
have become a paradigm for complex dynamical systems in general \cite{jensen}
(see also references in \cite{claussen2004}).

\mysection{Definition of ECA}
%
An elementary cellular automaton consists of an infinite
one-dimensional lattice of cells being either black (1) or white
(0), and a deterministic update rule. At each discrete time step,
a cell is updated $x_n^{t} \rightarrow x_n^{t+1}$ according to the
state of the next-neighbor sites and its own state one time step
before:
\begin{eqnarray}
x_n^{t+1}=f(x_{n+1}^t, x_n^t, x_{n-1}^t)
\end{eqnarray}
where $f$ (the rule) is determined by $8$ bits being the output of the possible
input bits $000$, $001$, ..., $111$.
As a consequence, there are 256 (ECA-)rules that are named rule 0 - 255.
In this article we focus on rules 90 and 150 defined by
\begin{eqnarray}
x_n^{t+1}=[x_{n-1}^t + r x_n^t + x_{n+1}^t] ~\mod~2
\end{eqnarray}
where $r=0$ defines rule 90, and $r=1$ rule 150, respectively.
As demonstrated earlier, rule 90 can be interpreted in the context of catalytic processes.
A process (catalysis) is initiated (or continued) when exactly 1
neighbor site is active whereas the process (catalysis) is stopped when too many, i.e. 2, or to less, i.e. no, neighbor sites are active \cite{claussen2004}.

\indent
A similar interpretation may be given for rule 150.
Catalysis at $x_n^t$ is stopped when no or two neighbor sites (now $x_n^t$ included)
are active and it is initiated (or continued) when one or three sites are active.
Note that both rules mimic local self-limiting reaction processes
\cite{dress84plath,%
otterstedt98}.

\begin{figure}[htbp]
\noindent
\epsfig{file=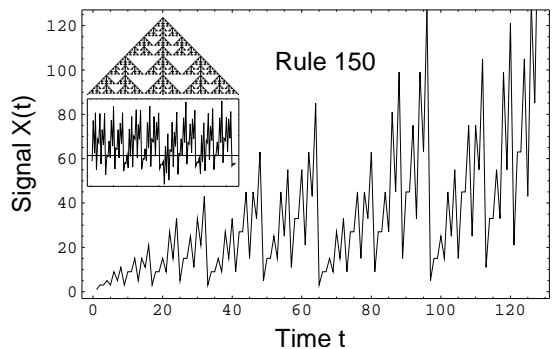,angle=0,height=4.5cm}               
\noindent
\caption{\label{fig_rule150_xt}
The first 128 time steps of the time signal $X(t)$ generated by rule 150.
Upper Inset:
Self-similar structure generated by rule 150 for the first 64 time steps.
Lower Inset: Normalized signal $Z(t)$; the straight line indicates $Z=0$.
}
\end{figure}

\begin{figure}[htbp]
\noindent
\epsfig{file=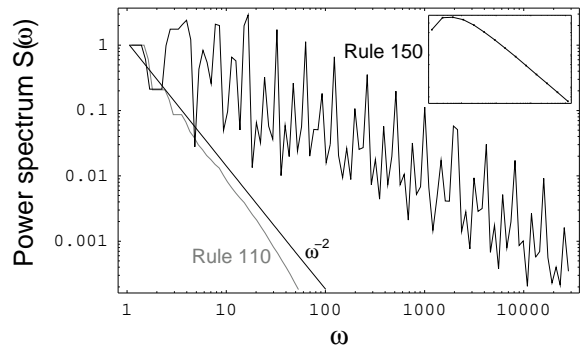,angle=0,height=4.5cm}               
\caption{\label{fig_rule150_spectrum}
Rule 150 and rule 110:
Averaged power spectrum of $Y(t)$ up to $T/8$ for $T=2^{18}$
using (incommensurable) $1.1^k$-bins, i.e.~
the $k$-th interval is  defined  by $[\lceil 1.1^k\rceil,\lceil
1.1^{k+1} \rceil]$
where the brackets $\lceil ~ \rceil$ denote upwards rounded integer values
(ceiling function).
The inset shows the rule 150 spectrum,
averaged using $2^k$-bins, i.e.{}
the $k$-th interval is  defined  by $[2^k,2^{k+1}-1]$.
Both averages correspond to a constant $\delta\omega/\omega$ ratio.
The graphs are well fitted by a power law with exponent $\alpha=1.27$.
The thermal $1/f^2$ decay of rule 110 (grey) 
as a typical member of Class 4
is shown for comparison.
}
\end{figure}

\mysection{Spectra of sum signals}
%
It is known that ECA on finite lattices for various boundary conditions
display no \oofs{} \cite{wolfram}. Rather than evaluating the rules
on finite lattices we calculate the evolution on an infinite lattice.
More precisely, we focus on a sum signal defined as the total {\em (in)activity, magnetization, etc.} of the whole system:
\begin{eqnarray}
X(t)=\sum_n x_n^t.
\end{eqnarray}
We have systematically investigated all 256 rules, for 
localized initial conditions (i.e. single 1, 11, 101, 111, ...), as follows.
The sum signal for non-trivial rules
exhibits increasing mean $\langle X\rangle_{t}$ well fitted by a power law in time
\footnote{In \cite{claussen2004} we have shown this both numerically and analytically
for rule 90. For other rules it is also easy to derive analytically.}.
Consequently, we focus on the detrended sum signal defined by
\begin{eqnarray}\label{yt}
Y(t)=X(t) - f(t)
\end{eqnarray}
where the coefficients of $f(t)=a t^b$ are fitted.
However for some ECA $Y(t)$ possesses an increasing mean variance.
Thus we investigate for each automaton another signal (and its
spectrum)
\begin{eqnarray}\label{zt}
Z(t)=Y(t)/\langle Y\rangle_{\{t-l+1,t+l\}}^{1/2}
\end{eqnarray}
where $2l$ is the width of a sliding window that normalizes the
fluctuations of the detrended signal $Y(t)$ according to the
method of detrended fluctuation analysis (DFA) applied for
non-equilibrium processes \cite{hu2001}.
We have calculated the corresponding power spectra
$|X(\omega)|^2$, $|Y(\omega)|^2$ and $|Z(\omega)|^2$ for all 256
ECA. It turns out that that 25 of the 256 rules exhibit \oofs{}
whereas 231 do not (see table \ref{table}). 23 of those automata that exhibit \oofs{}
display Sierpinski patterns, i.e. well studied self-similar
structures \cite{claussen2004}. Their spectra are extensively
investigated
in \cite{claussen2004} exhibiting \oofs{} with exponents $1.15\pm0.05$.\\
\begin{table}
\begin{tabular}{c|c}
Class & ECA rule number  \\ \hline
1 &  218 \\ \hline
2 &  (26, 82, 167, 181), (154, 210) \\ \hline
3 &  (18, 183), (22, 151), (60, 102, 153, 195), (90, 165), \\
  &  (122, 161), (126, 129), (146, 182), {\bf 105}, {\bf 150} \\ \hline
4 &  - \\
\end{tabular}
\caption{\label{table} Rules that produce $1/f^\alpha$ spectra.
Rules in brackets belong to one equivalence class. Rules 105 and
150 (bold) produce spectra with power law exponents about
$\alpha=1.3$. All other listed rules exhibit spectra with
exponents about $\alpha=1.2$. The 231 rules not listed are not
capable to produce $1/f^\alpha$ spectra,
e.g. most of the spectra display no power law decay,
or exhibit {\em thermal} $1/f^2$ spectra
(see Fig.~\ref{fig_rule150_spectrum}).
 }
\end{table}
\indent The two other rules, i.e. 105 and 150, show a different
behavior. Here we focus on rule 150 \footnote{Rule 105 is simply
the inverse of rule 150, i.e.
$f_{\text{105}}(a,b,c)=1-f_{\text{150}}(a,b,c)$.}. The first 128
time steps of the evolution for a single 1 as initial condition is
depicted in 
Fig.~\ref{fig_rule150_xt} (upper inset).
%
%
It is a Sierpinski-like self-similar structure. However the fractal dimension
differs from the Sierpinski gasket ($d=1.58$)
being the golden mean $d=(1+\sqrt{5})/2\approx 1.69$.
Fig.~\ref{fig_rule150_xt} shows also  the corresponding signals $X(t)$ and $Z(t)$.
The spectrum $Y(\omega)$ is displayed in Fig.~\ref{fig_rule150_spectrum}.
For $\omega$ not too small, the averaged spectrum
exhibits a straight line in the log-log-plot verifying a power law behavior.
Depending on the average process and fit range we obtain a fit exponent
of about $\alpha=1.27\pm0.05$.
Due to dominating randomness, members of classes 3 and 4
typically produce thermal $1/f^2$ spectra (see Fig.
\ref{fig_rule150_spectrum}).
%

\begin{figure}[htbp]
\noindent
\epsfig{file=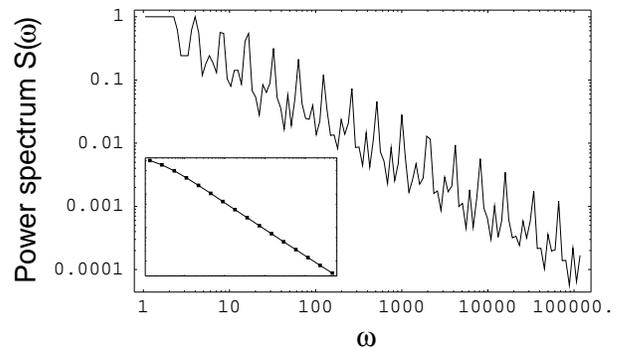,angle=0,height=4.5cm}               
\caption{\label{fig_gen1_spectrum}
Averaged power spectrum of the detrended signal (\ref{signals})
for $\delta=1.31184$ up to $T/8,~T=2^{20}$
using (incommensurable) $1.1^k$-bins, i.e.{} the $k$-th interval is  defined  by $[\lceil 1.1^k\rceil,\lceil 1.1^{k+1} \rceil]$
where the brackets $\lceil ~ \rceil$ denote upwards rounded integer values
(ceiling function).
The inset shows the spectrum,
averaged using $2^k$-bins, i.e.{} the $k$-th interval is  defined  by $[2^k,2^{k+1}-1]$.
Both correspond to a constant $\delta\omega/\omega$ ratio possessing fit exponents of about
$\alpha=0.93$. The corresponding spectrum for the variance detrended signal exhibits power spectra around $\alpha=1.0$. The theoretical exponent is $\alpha=1.0$.
}
\end{figure}

\mysection{Fractal signals produce \oofs}
%
All ECA that are capable to produce a self-similar structure exhibit \oofs.
Hence one may naively expect that every
(self-similar) fractal structure produces \oofs.
However it is important to note that this is not the case.
There are many fractals like the Koch snow flake, Cantor dust etc.
exhibiting no \oofs\  because their respective sum signals simply grow
exponentially \cite{mandelbrot}.\\
\indent Rather than a geometric approach we focus on fractal {\em
signals} itself. Thus we now generalize the recently investigated
Sierpinski signal \cite{claussen2004}. As we will see, the
generalized signal is capable to model \oofs{} producing spectra
with continuously tunable power law exponents.
More precisely, we consider the signal
\begin{eqnarray}\label{signals}
X_{\delta}(t)= 2^{\delta \sum_j \sigma_j\{t\}}
\end{eqnarray}
where $\sigma_j$ is the $j$th bit of the binary decomposition of the discrete time $t=0,1,2,...$.
For $\delta=1$ we have shown recently both numerically and analytically that the signal
exhibits {\oofs} with $\alpha$ close to unity.
The special ansatz, eq. (\ref{signals}), represents a straightforward
generalization of the closed form for the sum signal of the Sierpinski pattern produced by
rule 90 \cite{claussen2004}.
In the next paragraph we show that for deviations from $\delta=1$ the signal
can produce {\oofs} within a wide range of exponents $\alpha$.\\
\indent
In analogy to the calculation in Ref. \cite{claussen2004},
we calculate the periodogram $X(\omega)$
of the time signal (\ref{signals}) analytically:
\begin{eqnarray}\label{xom}
X(\omega)&=& \sum_{t=0}^{2^N-1} {\rm{}e}^{{\rm{}i}\omega t} X_\delta(t)\nonumber\\
         &=& \sum_{\{\sigma_0,\ldots,\sigma_{N-1}\}} \exp ({\rm{}i}\omega \sum_j \sigma_j 2^j)
~X_\delta( \sum_j\sigma_j 2^j ) \nonumber\\
&=& \sum_{\{\sigma_0,\ldots,\sigma_{N-1}\}} \prod_{j=0}^{N-1}
   \exp\left(  \sigma_j ({\rm{}i} \omega  2^j + \delta\ln 2) \right) \nonumber\\
&=& \prod_{j=0}^{N-1} \sum_{\{\sigma_j\}} \exp\left(  \sigma_j
   ({\rm{}i} \omega  2^j + \delta\ln 2) \right) \nonumber\\
&=& \prod_{j=0}^{N-1}
   \left( 1+ \exp ({\rm{}i} \omega  2^j + \delta\ln 2) \right).
\end{eqnarray}
The absolute value of $X(\omega)$ simplifies
to a trigonometric product which the logarithm converts into a sum:
\begin{eqnarray}
\ln |X(\omega)|^2 = \sum_{j=0}^{N-1} \ln [1+2^{2\delta}+2^{1+\delta} \cos(\omega 2^j)].
\label{n1}
\end{eqnarray}
We roughly estimate the sum in eq.~(\ref{n1}) replacing
the sum by an integral, and substituting $y=\omega 2^j$,
\begin{eqnarray}
\ln |X(\omega)|^2 & \approx& \int_{0}^{N-1} \ln [1+2^{2\delta}+2^{1+\delta} \cos(\omega 2^j)]
{\rm d}j
\label{n2}
\\
\label{bothints}
&=& \int_{\omega}^{\omega 2^{N-1}} \frac{\ln [1+2^{2\delta}+2^{1+\delta}\cos y]}{y\ln 2} {\rm d}y.
\end{eqnarray}
As $\ln(a+bx) \approx \ln(a)
+\frac{b}{a}x$ for $|x|\ll 1$,
we obtain
\begin{eqnarray}
\nonumber
\ln |X(\omega)|^2 \approx \frac{\ln (1+2^{2\delta})}{\ln 2}
\!
\int_{\omega}^{\omega
2^{N-1}}
\!\!
 \frac{ {\rm d}y}{y} +\\
                           \frac{2^{1+\delta}}{(1+2^{2\delta})\ln 2}
\!
\int_{\omega}^{\omega
2^{N-1}}
\!\!
               \frac{\cos(y)}{y} {\rm d}y.
\end{eqnarray}
\normalsize
\noindent
The integral over the integral cosine
is nearly
independent
of the upper boundary for high values of the boundary. Thus, we
can substitute the upper boundary $\omega 2^{N-1}$ by some
$N$-dependent constant, say $c_N\gg 1$. Finally, replacing the
cosine by one yields immediately a rough approximation of the power spectrum:
\begin{eqnarray}\label{result}
|X(\omega)|^2  \approx c_N'
\omega^{-
\frac{2^{1+\delta}}{(1+2^{2\delta})\ln 2}}.
\end{eqnarray}
For a given power law exponent $0<\alpha\le 1/\ln 2\approx 1.44$,
we obtain $\delta$ from eq.~(\ref{result}) as
\begin{equation}
\delta = \frac{\ln\Big(\frac{1+\sqrt{1-\alpha^2 (\ln 2)^2}}{\alpha\ln 2}\Big)}{\ln 2}.
\label{deltafromalpha}
\end{equation}
To generate signals with goal exponents, e.g.,\
$\alpha_1=0.8, \alpha_2=1.0, \alpha_3=1.2$,
one can use the corresponding value of
$\delta$ according to Eq.~(\ref{deltafromalpha}).
Fig.~\ref{fig_gen1_spectrum} shows the spectrum
of the detrended signal (\ref{signals})
for $\delta_2=1.31184$ (corresponding to $\alpha_2=1.0$).
For $\delta_1=1.72425$
and $\delta_3=0.902749$
the  individual spectra exhibit similar graphs (not shown).
Depending on the averaging the power law fits
giving $\alpha_1=0.8\pm 0.1$, $\alpha_1=0.95\pm 0.05$ and $\alpha_3=1.2\pm0.05$,
are in good agreement with the theoretical results.

\mysection{Two-dimensional automaton}
%
While one-dimensional experimental setups as in
\cite{otterstedt98} seem to be quite artificial for
(self-limiting) catalytic processes, two-dimensional dynamics is
more generic.
Consider the Sierpinski dynamics on a $(i,j)$-plane:
\begin{eqnarray}
x_{i,j}^{t+1}=[x_{i+1,j}^t+x_{i-1,j}^t+x_{i,j+1}^t+x_{i,j-1}^t ] ~\mod~2.
\end{eqnarray}
For a single 1 as initial condition
on a plane the sum signal $X_{2D}(t)=\sum_{i,j} x_{i,j}^t$
generates the sequence
\begin{eqnarray}\label{x2d}
X_{2D}(t)=1, 4, 4, 16, 4, 16, 16, 64, \ldots
\end{eqnarray}
More precisely, the recurrence relation generating eq.\ (\ref{x2d}) is given by
$u_n \rightarrow u_{n+1}=(u_n, 4 u_n)$ for $u_0=(1)$.\\
\indent
First, if the factor 4 is replaced by 2, the relation becomes
equivalent to the 1d-Sierpinski signal $X_{1D}(t)$ in Ref.\ \cite{claussen2004}.
Second, we obtain $X_{2D}(t)=X_{1D}(t)^2$ and therefore $X_{2D}(t)=X_{\delta=2}(t)$.
Thus the generalized Sierpinski pattern in two dimensions exhibits
\oofs{} with exponents around the value according to eq.\
(\ref{result}) for $\delta=2$, that is $\alpha=0.679$. We
numerically verified the value obtaining exponents around
$\alpha=0.7$ as expected.

\mysection{Conclusions}
%
Elementary Cellular automata are a paradigm for
emergence of complex spatiotemporal
behavior from extremely simple dynamics.
We systematically investigated all 256 elementary cellular automata.
As expected, among those as (nested) periodic/chaotic classified rules (classes 2 and 3)
there are various rules that display $1/f^\alpha$ spectra (see table \ref{table}).
Unexpectedly, on the one hand
all rules classified as complex display {\em no} $1/f^\alpha$ spectra,
while on the other hand, the {\em trivial} rule 218 does
(being a member of class 1).
It is important to note that the numerically calculated spectra are robust against noise, that is,
the fit exponents change only slightly for other initial conditions
than a single seed.\\
\indent
Moreover we generalized the approach of a sum signal
introduced in \cite{claussen2004}
to derive analytically the spectra of the 2D Sierpinski automaton.
The investigated fractal signals (\ref{signals})
serve also as a fit model for signals produced by elementary cellular automata rules.
We have obtained a time series generator with continuously tunable power law decay exponent.
The tailored signals represent analytically tractable (nontrivial)
$1/f^{\alpha}$ generators that shed light on
the arcane mechanisms of $1/f^{\alpha}$ spectra.\\
\indent 
From our results, we expect that in experimental systems
showing spatiotemporal pattern formation similar to the ECA patterns,
the power spectra of the total (in)activity will exhibit 
power law behavior within a certain range.
\vspace*{-1ex}
%


\end{document}